\documentclass[conference]{IEEEtran}
\IEEEoverridecommandlockouts
\usepackage{cite}
\usepackage{amsmath,amssymb,amsfonts}
\usepackage{algorithmic}
\usepackage{graphicx}
\usepackage{textcomp}
\usepackage{xcolor}
\usepackage{hyperref}
\usepackage{todonotes}

\usepackage[acronym,toc,symbols]{glossaries}
\glsdisablehyper
\newacronym{iot}{IoT}{Internet of Things}
\newacronym{api}{API}{Application Programming Interface}
\newacronym{nef}{NEF}{Network Exposure Function}
\newacronym{ue}{UE}{User Equipment}
\newacronym{k8s}{K8s}{Kubernetes}
\newacronym{qos}{QoS}{Quality of Service}
\newacronym{cni}{CNI}{Container Networking Interface}
\newacronym{xr}{XR}{eXtended Reality}
\newacronym{vxlan}{VXLAN}{Virtual Extensible LAN}
\newacronym{ar}{AR}{Augmented Reality}
\newacronym{vr}{VR}{Virtual Reality}
\newacronym{slam}{SLAM}{Simultaneous Localization and Mapping}
\newacronym{pdu}{PDU}{Protocol Data Unit}
\newacronym{ran}{RAN}{Radio Access Network}
\newacronym{5qi}{5QI}{5G QoS Identifier}
\newacronym{qfi}{QFI}{QoS Flow Identifier}
\newacronym{tc}{TC}{Traffic Control}
\newacronym{sdn}{SDN}{Software-Defined Networking}
\newacronym{tos}{ToS}{Type of Service}
\newacronym{amf}{AMF}{Access and Mobility Management Function}
\newacronym{seal}{SEAL}{Service Enabler Architecture Layer}
\newacronym{gnb}{gNB}{Next Generation NodeB}
\newacronym{upf}{UPF}{User Plane Function}
\newacronym{ape}{APE}{Absolute Position Error}
\newacronym{fwmark}{fwmark}{firewall mark}
\newacronym{mec}{MEC}{Multi-access Edge Computing}
\newacronym{htb}{HTB}{Hierarchical Token Bucket}

\newcommand\copyrighttext{%
  \footnotesize © 2024 IEEE.  Personal use of this material is permitted.  Permission from IEEE must be obtained for all other uses, in any current or future media, including reprinting/republishing this material for advertising or promotional purposes, creating new collective works, for resale or redistribution to servers or lists, or reuse of any copyrighted component of this work in other works.}
\newcommand\copyrightnotice{%
\begin{tikzpicture}[overlay, remember picture]
\node[anchor=south,yshift=10pt] at (current page.south) {\fbox{\parbox{\dimexpr\textwidth-\fboxsep-\fboxrule\relax}{\copyrighttext}}};
\end{tikzpicture}%
}

\def\BibTeX{{\rm B\kern-.05em{\sc i\kern-.025em b}\kern-.08em
    T\kern-.1667em\lower.7ex\hbox{E}\kern-.125emX}}
\begin{document}

\title{Enabling 5G QoS configuration capabilities for IoT applications on container orchestration platform
}

\author{\IEEEauthorblockN{1\textsuperscript{st} Yu Liu}
\IEEEauthorblockA{\textit{Ericsson Research} \\
Stockholm, Sweden \\
yu.a.liu@ericsson.com}
\and
\IEEEauthorblockN{2\textsuperscript{nd}  Aitor Hernandez Herranz}
\IEEEauthorblockA{\textit{Ericsson Research} \\
Stockholm, Sweden \\
aitor.hernandez.herranz@ericsson.com}
}

\maketitle
\begin{abstract}

Container orchestration platform is the foundation of modern cloud infrastructure. In recent years, container orchestration platform has been evolving to cross the boundary of device, edge, and cloud. More and more \gls{iot} applications such as robotics and \gls{xr} have been deployed across the device-cloud continuum through the container orchestration platform, e.g., the \gls{k8s} framework. Meanwhile, the rapid expansion of advanced communication technologies like 5G has endorsed the revolution in \gls{iot} applications as more network resource is available for critical \gls{iot} use cases.
This paper aims to integrate network configuration capabilities provided by a 5G \gls{nef} into the \gls{k8s} framework which is used to simplify application deployment in an orchestration in the device-cloud continuum. Specifically, a Linux fwmark-based network \gls{qos} configuration method is proposed to expose the \gls{qos} information from an overlay network that is used by the container orchestration platform to the underlay network. A \gls{cni} plugin-based implementation is demonstrated to perform \gls{qos} configuration for the 5G network. The proposed solution is validated with an existing localization and mapping application to verify the feasibility.
The proposed solution has the following benefits:
(1) The solution is a Kubernetes-native approach which adopts the \gls{cni} plugin mechanism.
(2) The solution can expose the \gls{qos} information from an overlay network to an underlay network in a non-intrusive manner.
(3) No packet manipulation is required to greatly reduce the overhead for packet processing.
(4) It extends the  \gls{k8s} bandwidth limit feature from on-node to the access network.
(5) It is compatible with the 5G infrastructure without any alteration or adding extra complexity.

\end{abstract}

\begin{IEEEkeywords}
    5G, QoS, container orchestration platform, container network interface, device-cloud continuum
\end{IEEEkeywords}

\glsresetall
\section{Introduction}
\label{sec:intro}
\copyrightnotice
A container orchestration platform such as \gls{k8s} or K3s usually manages a cluster of nodes that span different networks. Considering the network heterogeneity and to simplify communication within the cluster, an overlay network is usually established to enable communication among application pods. Typical networking addons such as flannel and calico can create overlay networks using \gls{vxlan}, IP in IP protocols or secured channels using IPSec or WireGuard.

The integration of cloud computing and 5G technology has given rise to the transformative concept of \gls{mec}. The merge promises higher computing capabilities and lower network latency at the network edge, which largely enables applications such as \gls{ar}/\gls{vr} and leads to new deployment paradigm such as cloud robotics~\cite{cloud_robotics_survey}. One of the key enablers lies in that an access network (e.g., 5G, and WiFi, etc.) can provide different \gls{qos} to applications catering to different network traffic priority demands to satisfy application performance requirements while distributing network resource in an efficient manner. For instance, a \gls{ue} can establish multiple \gls{qos} flows in the same \gls{pdu} session to a 5G network.

The integration of 5G \gls{qos} configuration capabilities to the existing K8s ecosystem becomes a fundamental and demanding feature that can benefit \gls{iot} applications but remains challenging. Due to the use of an overlay network in K8s, application-specific QoS information that are hidden in the overlay network packets can hardly be extracted in the underlay network without extra overhead.
Therefore, the goals for this study is to investigate options of integrating network exposure \gls{api}s provided by a 5G network (e.g. \gls{nef} or \gls{seal}) into the lightweight Kubernetes which is used to simplify application deployment in an orchestration in the device-cloud continuum, to implement a bridge between container-orchestration and network \gls{qos} configuration. The platform-network interaction is validated with an existing IoT application as well as the potential impact on the application.

This paper is organized as follows. In Section~\ref{sec:sota} the background of \gls{qos} configuration in cellular networks and in the container orchestration platform is introduced and the pitfalls in existing solutions are pointed out. In Section~\ref{sec:solution}, we introduce the \gls{cni} plugin-based solution to enable \gls{qos} configuration for container orchestration platform as well as theoretical fundamentals. In Section~\ref{sec:experiments}, an experimental implementation of the traffic priority \gls{cni} plugin is detailed and verified by applying to an \gls{slam} application. Section~\ref{sec:conclusions} concludes the paper.

\section{State of the art}
\label{sec:sota}

\subsection{QoS configuration on cellular networks}

The advent of new cellular access networks has enabled a host of use cases that enhance quality of life and sustainability. These networks, particularly 5G, offer a reliable solution for interconnecting machines and humans, eliminating the need for wired connections.

In many of the use cases, it is desirable to allow the devices, i.e., \gls{ue}, to be able to control the traffic and apply different \gls{qos} to different applications or flows running locally in the \gls{ue}. This control allows the network to treat different types of traffic differently and provide specific resources as required. For example, in a 5G network, service data from the \gls{ue} can be classified into different flows marked by different \gls{qfi}~\cite{ts23.501}, which allows the \gls{ue} traffic to have different \gls{qos} configurations and receive different treatments when traversing the 5G network. \glspl{5qi}~\cite{ts23.501} are used to define a set of characteristics of the flow in terms of flow bit rate resource type, priority levels, packet delay budget and packet error rate, averaging window, and maximum data burst volume, among others.

There are several methods to configure \gls{qos} in a \gls{ran} such as 5G. The most common means is to configure the \gls{qos} through  \gls{nef}. \gls{nef} allows applications either running on a \gls{ue} or external data networks to request a \gls{qos} for a \gls{pdu} session that is to be established or request a \gls{qos} change for a given \gls{pdu} session, via the \glspl{api} exposed by \gls{nef}. Alternatively, a \gls{ue} can also actively initiate a \gls{qos} configuration or modification for a given \gls{pdu} session through interactions with \gls{amf}. The technical details of \gls{qos} configuration in a 5G network are out of the scope of this paper and can be found in \cite{ts23.501}.

\subsection{Network QoS on container orchestration}
The K8s orchestration platform utilizes a \gls{cni} plugin mechanism to set up the network stack for containers. Existing network \gls{qos} configuration in the \gls{cni}~\cite{cni} is provided by the bandwidth plugin~\cite{bandwidth_plugin}, which configures a Token Bucket Filter (TBF) queuing discipline (qdisc) on both ingress and egress traffic which results in the traffic being shaped. However, the bandwidth plugin can only shape local traffic that takes place within a node.

Other mechanisms for traffic shaping and configuration based on \gls{cni} plugins and network plugins for \gls{k8s} have been developed in relation to \gls{sdn}. Examples include OpenShift SDN plugins such as Cisco ACI SDN, Flannel SDN, NSX-T SDN, Nuage SDN, and Kuryr SDN~\cite{openshift}. Yet, these \gls{sdn} based approaches are typically enforced within nodes but not applied to the communication network.

Another project serving the traffic differentiation purpose is NBWguard~\cite{nbwguard}, which is proposed by the research community. It provides network \gls{qos} in the same way CPU or memory are limited in \gls{k8s} via the control groups (cgroups),  a Linux kernel feature to limit and prioritize resources~\cite{man7}. The three \gls{qos} modes defined in \gls{k8s}, i.e., \textit{BestEffort, Burstable and Guaranteed}, are extended to network in this work. Internally, NBWguard also uses \gls{tc} and applies \gls{htb} queuing discipline (qdisc). However, similar to the bandwidth plugin, NBWguard's scope is limited to local traffic shaping on the node.

Another approach to implement network functions is through the native device plugin feature enabled by the \gls{k8s} framework. One example is the SR-IOV network device plugin~\cite{sriov}. This approach enables a pod to interact with physical network devices to add customized processing logic to workloads. Nevertheless, this device plugin-based solution is hardware-dependent and can hardly be regarded as a generic solution.

\begin{figure}[tbp]
    \centering
    \includegraphics[width=\linewidth]{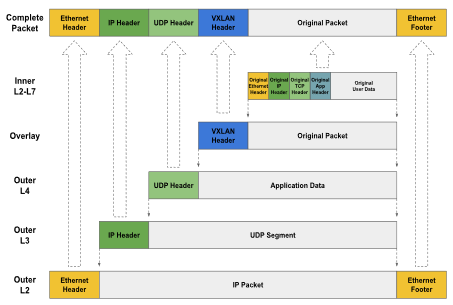}
    \caption{Anatomy of VXLAN network packet. \\Source: https://projectcalico.docs.tigera.io/about/about-networking}
    \label{VXLAN}
    
\end{figure}

\begin{figure}[t]
    \centering
    \includegraphics[width=\linewidth]{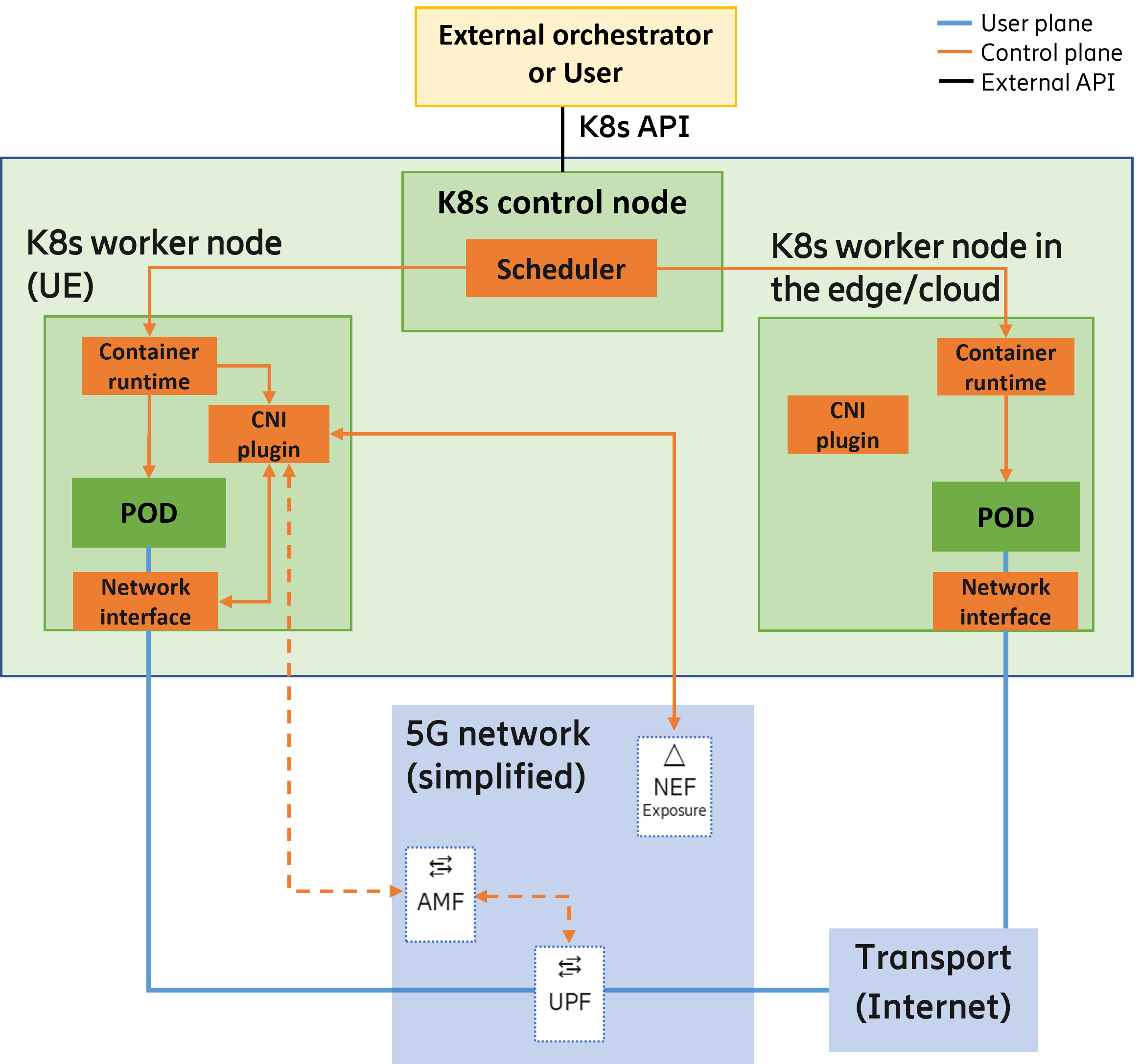}
    \caption{Architecture of the proposed solution. It depicts a UE application running on a K8s pod is communicating with an edge/cloud service through the 5G network. A traffic priority CNI plugin is implemented in the UE to interact with the 5G network through either AMF or NEF to configure the application pod's QoS. Note that the network interface may represent a physical interface or a 5G network modem stack or modem manager.}
    \label{architecture}
    \vspace{-0.5cm}
\end{figure}

\subsection{Problems with existing solutions}
To enable a seamless integration of 5G \gls{qos} configuration into the container orchestration platform like K8s, the \gls{qos} demanded by the application running on the K8s platform must be delivered to the 5G network. However, many challenges remain here.
\subsubsection{Overlay networks}
An overlay network can simplify addressing and routing between application pods within a cluster. However, due to the nature of an overlay network that IP packets generated by an application pod are encapsulated inside an underlay network packet, which indicates the \gls{qos} or \gls{tos} information that are hidden in the overlay IP packets are invisible to underlay network interface or network filters unless a decapsulation is performed on the packets, as illustrated in Figure~\ref{VXLAN}. Also, this would require manual manipulation on the overlay packets including inserting \gls{qos} information to the overlay packet header or payload. This manipulation, decapsulation and re-encapsulation procedure can greatly increase the overhead and reduce the network communication performance.

\subsubsection{Limitation of traffic prioritization in K8s}
In \gls{k8s}, the network capability is enabled by a plethora of \gls{cni} plugins which are highly environment-dependent to address the dynamic needs of networking in different scenarios. Therefore, the \gls{cni} specification is well defined for a pluggable network solution. Among existing \gls{cni} plugins, most are catering to the connectivity challenges while the traffic prioritization and characteristics are still with limited support. E.g., the native bandwidth plugin provides an approach to limit ingress and egress bandwidth, though the effect is restricted at the node. Other plugins such as Calico features network policy configurations which can be utilized to enable or disable an application pod to communicate with various network entities. However, more complicated configuration capabilities that can expose pod-specific \gls{qos} to underlay networks and enable \gls{qos} configuration in access networks are still missing.
To our knowledge, no existing solutions are available to provide a similar QoS configuration capability as discussed in this paper.

\section{Proposed solution}
\label{sec:solution}

\subsection{Architecture}
Figure \ref{architecture} illustrates the architecture of the proposed solution that can expose the 5G network \gls{qos} configuration capability to containers running on top of K8s at a \gls{ue} through the standard \gls{cni} plugin approach.

In this architecture, the network \gls{qos} requirements can be inserted into the K8s control plane either by an end user or by an external orchestrator through the \textit{kube-apiserver} component using the standard K8s API. Upon receiving the configuration, K8s will extract the network \gls{qos} information and schedule the pod creation procedure. This procedure is initiated by the \textit{kubelet} component that is residing in every worker node, and handled by the container runtime. Among multiple tasks, the container runtime would specifically call a series of  \gls{cni} plugins according to the predefined  \gls{cni} network configuration, to setup the network environment for the created pod.

\begin{figure}[t]
    \centering
    \includegraphics[width=\linewidth]{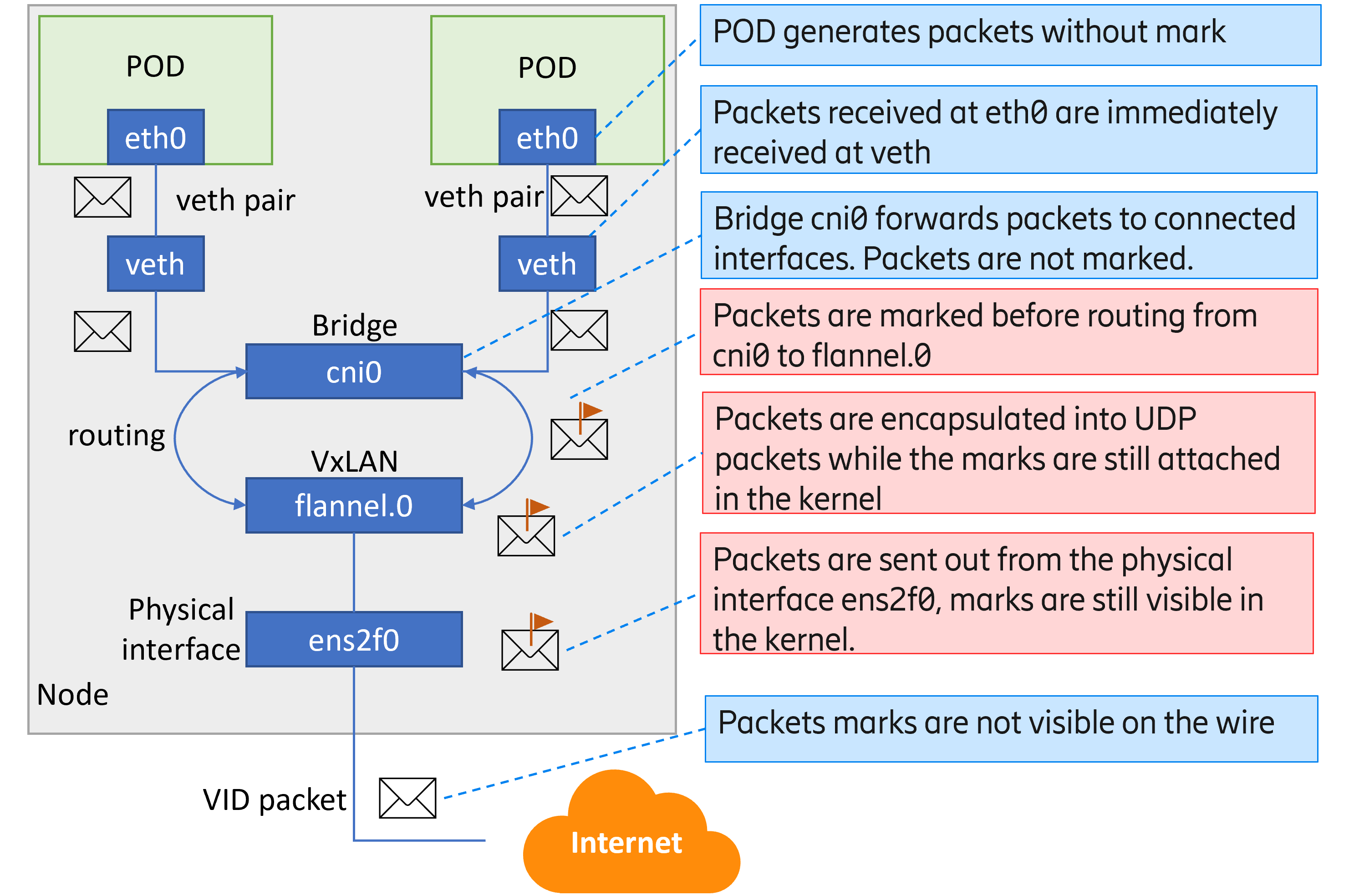}
    \caption{Linux packet fwmark visibility in a typical Kubernetes networking stack.}
    \label{fwmark_visibility}
    \vspace{-0.5cm}
\end{figure}

Our solution leverages the \gls{cni} plugin mechanism by adding a traffic prioritization plugin in the CNI plugin chain to enable K8s to interact with the 5G stack. Once invoked, this plugin would setup an IP flow for all egress traffic from the pod and then interact with the 5G stack to establish a unique 5G \gls{qos} flow that can satisfy the demanded network \gls{qos}, either via the \gls{nef} component which is through the data plane or the \gls{amf} component that is directly requested through the control plane. Upon successful establishment of the \gls{qos} flow, the plugin can create a filter applied to the physical interface to redirect the IP flow associated with the pod to the 5G \gls{qos} flow. In this way can each pod be mapped to a unique 5G \gls{qos} flow in a native manner that is supported by the K8s infrastructure.

\subsection{Linux packet fwmark visibility and availability}
To accomplish the 5G \gls{qos} configuration for a specific pod, a key step is to distinguish the traffic from the pod from all the other traffics to create a unique IP flow. Considering the nature of a typical \gls{k8s} networking architecture in which VXLAN is utilized to enable pod-to-pod communication, one approach is to label the packets at the VXLAN interface. After the encapsulation of the overlay network packets, the encapsulated packets can be labelled by inserting labeling information in the packet header or the payload, which can be filtered to create an IP flow thereafter. However, this approach would greatly increase the overhead and inevitably add extra complexity to the VXLAN implementation. Therefore, a solution that can expose the pod \gls{qos} requirement that is hidden in the overlay network to the underlay network in a transparent and low footprint manner is in demand.

Among potential solutions, one promising approach is to leverage the Linux \gls{fwmark}. The fwmark is a 32 bits field that allows to tag a packet in the Linux kernel, which doesn't need to manipulate the packet itself. The fwmark field can be set by iptables and be used to classify packets into different IP flows using a TC filter, which is a Linux-native approach and does not introduce extra overhead to packet processing.

\subsubsection{Fwmark visibility}
Figure \ref{fwmark_visibility} shows the Linux fwmark visibility in a typical \gls{k8s} networking environment where flannel is used to create the VXLAN network. When a packet is generated by a pod, it is sent out through the eth0 interface and then immediately received by the veth interface that is a veth pair device. Since all veth devices are connected to a bridge device, i.e., cni0 in the example, the packet is forwarded to all devices connected to the bridge. Until now, the fwmark is not tagged. When the packet is sent from cni0 to the VXLAN interface flannel.0, a routing is needed. Before the routing procedure, a rule created in the iptables at the mangle table and prerouting chain can be applied to mark the fwmark field of the packet. When the packet is routed to flannel.0, it is encapsulated into a UDP packet as any regular VXLAN packets. After the encapsulation, the fwmark is still visible in the kernel, and the visibility is maintained till the packet is transmitted out of the host by the physical interface. This observation verifies the assumption that the fwmark can be utilized to expose \gls{qos} information from the overlay to the underlay network.

\subsubsection{Fwmark availability}
Another critical aspect of the fwmark approach is the availability of the bits that can be utilized in \gls{qos} mapping.

\begin{table}[ht!]
    \vspace{-0.5cm}
    \centering
    \caption{Linux fwmark registry for bitwise usage \cite{fw_registry}}
    \label{fwmark_availability}

    \begin{tabular}{|p{2cm}|p{2cm}|p{2cm}|}
        \hline
        \textbf{Bits} & \textbf{Mark mask} & \textbf{Software} \\ \hline
        0-12,16-31    & 0xFFFF1FFF         & Cilium            \\ \hline
        7             & 0x00000080         & AWS CNI           \\ \hline
        13            & 0x00002000         & CNI Portmap       \\ \hline
        14-15         & 0x0000C000         & Kubernetes        \\ \hline
        16-31         & 0xFFFF0000         & Calico            \\ \hline
        17-18         & 0x60000            & Weave Net         \\ \hline
        18-19         & 0xC0000            & Tailscale         \\ \hline
    \end{tabular}
\end{table}
Table \ref{fwmark_availability} demonstrates the bitwise usage of the fwmark by a series of software. It is noticed that in a \gls{k8s} environment where popular network plugins such as CNI portmap, Calico or Weave Net are installed, most bits from bit 13 to bit 31 are occupied. In an extreme case where the Cilium plugin is used, there are only 3 bits left for other purpose. Therefore, to adopt the fwmark for \gls{qos} mapping, potential conflicts with existing software must be considered.

\begin{figure}[tb]
    \centering
    \includegraphics[width=0.8\linewidth]{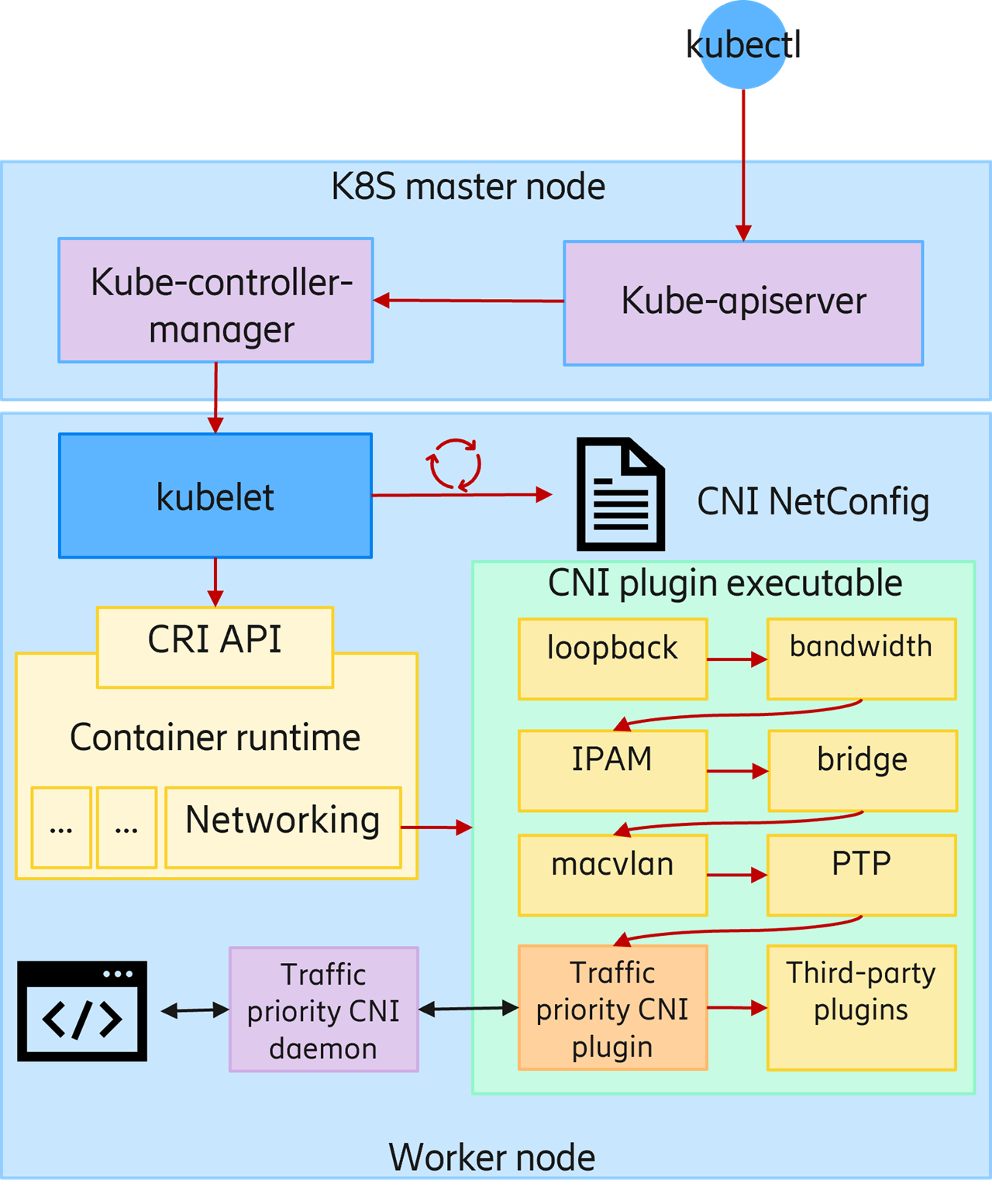}
    \caption{CNI plugin based \gls{qos} configuration. A traffic priority CNI plugin is invoked by kubelet in the CNI plugin chain. It interacts with the external access network, e.g., 5G network through a daemon to configure the QoS demanded by an application pod.}
    \label{cni_plugin}
    \vspace{-0.5cm}
\end{figure}
\subsection{CNI plugin based QoS configuration}
Figure \ref{cni_plugin} illustrates how a CNI plugin-based solution can be used to accomplish the \gls{qos} configuration for a given pod. Following the top-down order, the \gls{qos} information is passed into K8s via \textit{kube-apiserver}, which is then forwarded to \textit{kube-controller-manager} and delivered to \textit{kubelet} at the worker node where the pod is to be deployed. Through the container runtime interface, the \textit{kubelet} component is able to interact with the container runtime to prepare necessary resource such as sandbox and container namespace for container creation. Among many tasks, the container runtime needs to create the networking environment for the pod by invoking a chain of CNI plugins, including the traffic priority CNI plugin that implements the \gls{qos} based traffic prioritization.

According to CNI plugin specification~\cite{cni_spec}, the CNI plugins are invoked one after another by the container runtime. During each call, the parameters are passed into the CNI plugin via the STDIN and environment variables. The plugin configuration information that is stored at the CNI NetConfig file is periodically fetched by the container runtime and fed into each CNI plugin through STDIN while container-specific information such as container ID, namespace, and interface name are passed through environment variables.

The CNI plugin and CNI daemon configure \gls{qos} flows through interaction with the access network. Upon receipt of the request from the CNI plugin, the CNI daemon program will map the pod-demanded \gls{qos} requirement to the actual \gls{qos} flow characteristics such as the 5G \gls{qos} identifier (5QI) in particular. After that, the daemon program can either request the \gls{qos} flow to be established through the \gls{nef} or through the 5G modem interface, depending on implementation. Once successful, the daemon program can proceed further to generate a fwmark/\gls{qos} pair, configure the iptables to mark the packets sent from the pod, and create a \gls{tc} filter to redirect the marked packets to the established \gls{qos} flow.

\subsection{Benefits and limitations of the solution}
To summarize, the proposed solution has the following advantages:
\begin{itemize}
    \item The solution provides application pod the capability to configure traffic priority in an access network and maps the priority configuration to \gls{qos} of the access network with a CNI plugin, which is a native approach to be integrated into existing container orchestration platform such as \gls{k8s}.
    \item There is no need to change the packet header or payload to embed \gls{qos} information.
    \item It allows to transparently expose traffic prioritization tags/annotation from packets in the overlay network (pod) to the underlay network (e.g., VXLAN and IPIP), or even encrypted networks (e.g., IPSec).
    \item It enhances the bandwidth limit feature of \gls{k8s} CNI plugin by extending the bandwidth limit scope from the network stack on the node to both the node and the access network.
    \item The solution focuses on mapping pod \gls{qos} requirements to \gls{qos} flows on cellular networks, between \gls{ue} and \gls{gnb} (assuming the connectivity between \gls{gnb} and the core network is perfect), which is fully compatible with the existing 5G network infrastructure without introducing any alteration or adding extra complexity to the 3GPP standard.
\end{itemize}

Meanwhile, the proposed solution is also limited to the following aspects:
\begin{itemize}
    \item It only considers the outgoing traffic, i.e., egress traffic or uplink traffic from the pod running in the \gls{ue}.
    \item The traffic differentiation or \gls{qos} is only considered in the \gls{ran}, between the \gls{ue} and the 5G core (\gls{upf}). It does not guarantee packet \gls{qos} in the data network, e.g., the Internet, which is usually not under control.
\end{itemize}

\section{Experiments and Validation}
\label{sec:experiments}

In this section, we demonstrate the implementation of the proposed CNI plugin and validate the applicability of the plugin.

\subsection{Demonstrative implementation}

\subsubsection{Traffic priority CNI plugin}
According to the CNI specification~\cite{cni_spec}, four commands must be implemented for each CNI plugin, i.e., ADD, DEL, CHECK, and VERSION, which are passed into the plugin with the CNI\_COMMAND environment variable. The ADD command is executed when a container is added to the network or modifications are applied. The DEL command is used to remove a container from the network or un-apply the modifications. The CHECK command verifies the container's networking is as expected while the VERSION command returns a supported CNI version list.

In the implementation, the ADD and DEL commands are emphasized to validate the feasibility. When the ADD command is called, the traffic priority CNI plugin would generate new fwmark and insert iptables rules for the added POD, request the 5G stack to create radio link, \gls{pdu} session, and \gls{qos} flow using the inferred 5QI value, and then add IP filter to redirect marked packets to the \gls{qos} flow. When the DEL command is called, the plugin would delete the created iptables rules, the corresponding fwmark for the specific POD, the IP filter, as well as the 5G QoS flow.

\subsubsection{Interaction with 5G emulator}
\begin{figure*}[tb]
    \centering
    \includegraphics[width=0.85\linewidth]{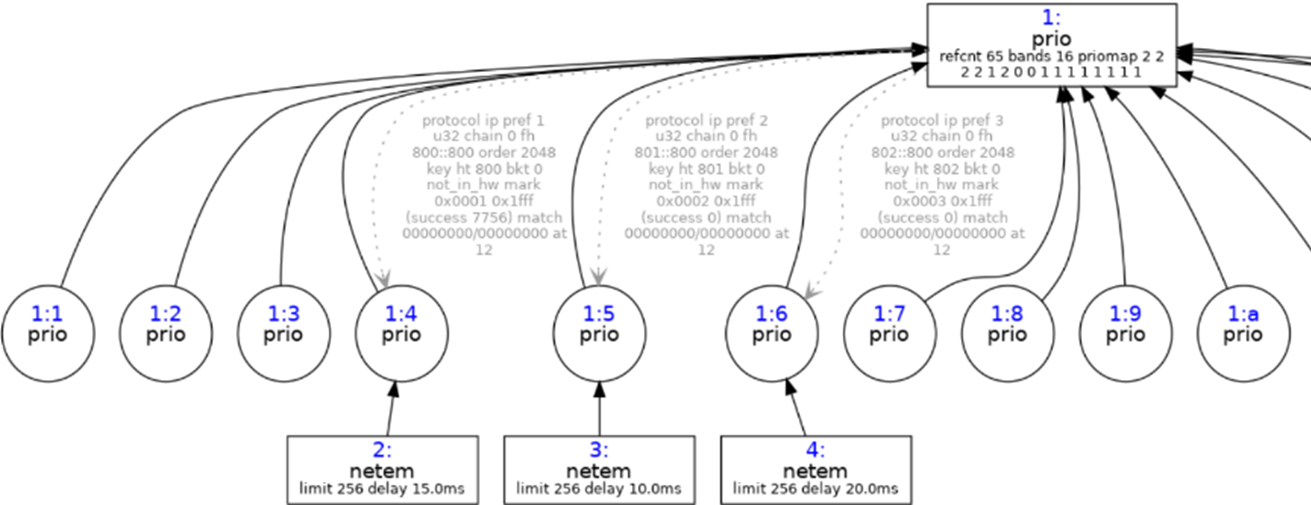}
    \caption{An example of the TC qdisc, class, and filter configurations after the \gls{qos} requirement is enforced by the 5G network emulator.}
    \label{tc}
    \vspace{-0.5cm}
\end{figure*}

A 5G network emulator built atop Linux \gls{tc} is used to emulate the 5G stack. It emulates the 5G network by configuring TC queuing disciplines, classes, and filters. Clients can interact with the 5G network emulator via exposed RESTful HTTP APIs to mimic the interaction with a real 5G network, which enables clients to create radio links, \gls{pdu} sessions, \gls{qos} flows and filters so as to realize traffic prioritization. Figure \ref{tc} shows an example of configured TC queuing disciplines, classes, and filters that classify traffic into different \gls{qos} flows. Specifically, three \gls{qos} flows corresponding to three different traffic priorities are highlighted, which are represented by network delays. Three filters are created to enqueue packets with unique fwmark values into the three \gls{qos} flows, respectively. In this way, the overlay network packets that are marked with a \gls{qos} fwmark can be redirected into different \gls{qos} flows in the 5G network.
\vspace{-4mm}

In the implementation, the CNI plugin can interact with the emulated 5G network and request the creation and deletion of a \gls{qos} flow through the provided HTTP API. \gls{qos} configurations requested by the CNI plugin are translated by the emulator into \gls{tc} commands which are applied to the physical network interface of the host node. In this way it emulates a UE pod to edge/cloud communication through the 5G network while different network \gls{qos} configurations can be requested.

\subsection{Validation: network QoS configuration for SLAM}
The proposed \gls{qos} configuration approach based on the CNI plugin is further validated in a real \gls{slam} application to verify the feasibility.
\subsubsection{SLAM testbed introduction}
The testbed used in the validation is built on top of the K3s framework. The underlying hardware include an Nvidia Jetson NX board (arm64) acting as a device node, an Nvidia Jetson AGX board (arm64) acting as an edge node, and other blade servers (amd64) acting as the cloud. The platform features multi-architecture support and low footprint, which enables applications to be deployed across device, edge, and cloud according to specific needs. Observability of application, platform, and hardware metrics is supported and visualized through Prometheus and Grafana that are deployed to the platform.

A \gls{slam} application, maplab\cite{maplab}, has been chosen for the case study as a distributed application. Execution of maplab consists of localization, mapping, and map optimization three phases, which can be run either in a centralized mode or distributed mode. The maplab application has been containerized with multi-architecture support and deployed on the K3s platform. 
The machine hall dataset\cite{Burri25012016} are utilized to run the \gls{slam} application.

\subsubsection{QoS configuration for SLAM}
Many researches have been conducted around the \gls{slam} application, e.g., in \cite{slam} the authors investigated the network condition's impact to \gls{slam} in distributed mode. As network latency increases, the \gls{slam} localization error also increases proportionally, which has been observed in all experimental datasets.

In the validation, the containerized \gls{slam} application is deployed to the aforementioned platform and configured to run in the high-offload distributed mode, i.e., the device node transmits both camera and inertial measurement unit data to the edge node and all computation tasks such as localization, mapping, and optimization take place at the edge.

Experiments are conducted in two categories namely \gls{qos}-unlimited and \gls{qos}-limited to compare the performance in respect of the \gls{ape} metric. For the \gls{qos}-unlimited case, no \gls{qos} is configured for the application pod while for the \gls{qos}-limited case, the pod running the device application is configured to use a \gls{qos} flow that has a 10 ms latency using the developed traffic priority CNI plugin. According to \cite{slam}, the localization error at 10 ms latency starts to be significant. The results are shown in Table \ref{comp1_table}, which shows the \gls{slam} application's performance of the \gls{qos}-limited category is slightly degraded in both the RMSE and Mean of \gls{ape} metrics compared to the \gls{qos}-unlimited category, which is in line with the expected results as shown in \cite{slam}. It also demonstrates the traffic priority CNI plugin is able to perform traffic prioritization of the network \gls{qos} for a given application deployed to the K8s platform so as to verify the feasibility.
\begin{table}[ht!]
    \vspace{-0.5cm}
    \centering
    \caption{Comparison of absolute position error (APE) between QoS-unlimited and QoS-limited SLAM experiments. The higher the worse.}
    \label{comp1_table}

    \begin{tabular}{|p{2cm}|p{2cm}|p{2cm}|}
        \hline
        \textbf{Category} & \textbf{RMSE (cm)} & \textbf{Mean (cm)} \\ \hline
        QoS-unlimited   & $9.24\pm0.14$        & $8.13\pm0.16$            \\ \hline
        QoS-limited     & $10.08\pm0.12$       & $8.85\pm0.11$           \\ \hline
    \end{tabular}
    \vspace{-0.5cm}
\end{table}

\section{Concluding remarks}
\label{sec:conclusions}
As an effort to facilitate the synergy of the cloud and the telecommunication domains, this study aims to expose the cellular network \gls{qos} configuration capability to applications running on a container orchestration platform such as K8s. A low footprint \gls{qos} mapping approach is proposed and implemented by leveraging the Linux fwmark feature and the K8s CNI plugin mechanism. The experimental validation shows the proposed approach can be applied to real \gls{slam} applications and perform prioritization of traffics that are configured with distinguished access network \gls{qos}. A quantitative performance evaluation of the solution can be conducted as a future study.  

In particular, the proposed solution addresses the challenge to expose overlay network packet's \gls{qos} information to the underlay network in a non-intrusive manner, which is significant to the container orchestration environment such as K8s where an overlay network is commonly used. The core essence of the solution can also be regarded as a complement to the existing 5G IP filter set-based approach that is used to configured 5G \gls{qos} flows defined in the 3GPP standard.

\bibliographystyle{IEEEtran}
\bibliography{ref}

\begin{thebibliography}{10}
\providecommand{\url}[1]{#1}
\csname url@samestyle\endcsname
\providecommand{\newblock}{\relax}
\providecommand{\bibinfo}[2]{#2}
\providecommand{\BIBentrySTDinterwordspacing}{\spaceskip=0pt\relax}
\providecommand{\BIBentryALTinterwordstretchfactor}{4}
\providecommand{\BIBentryALTinterwordspacing}{\spaceskip=\fontdimen2\font plus
\BIBentryALTinterwordstretchfactor\fontdimen3\font minus
  \fontdimen4\font\relax}
\providecommand{\BIBforeignlanguage}[2]{{%
\expandafter\ifx\csname l@#1\endcsname\relax
\typeout{** WARNING: IEEEtran.bst: No hyphenation pattern has been}%
\typeout{** loaded for the language `#1'. Using the pattern for}%
\typeout{** the default language instead.}%
\else
\language=\csname l@#1\endcsname
\fi
#2}}
\providecommand{\BIBdecl}{\relax}
\BIBdecl

\bibitem{cloud_robotics_survey}
M.~Afrin, J.~Jin, A.~Rahman, A.~Rahman, J.~Wan, and E.~Hossain, ``Resource
  allocation and service provisioning in multi-agent cloud robotics: A
  comprehensive survey,'' \emph{IEEE Communications Surveys \& Tutorials},
  vol.~23, no.~2, pp. 842--870, 2021.

\bibitem{ts23.501}
3GPP, \emph{3GPP TS 23.501 - System architecture for the 5G system (5GS)}.

\bibitem{cni}
\BIBentryALTinterwordspacing
``Container network interface.'' [Online]. Available:
  \url{https://www.cni.dev/plugins/current}
\BIBentrySTDinterwordspacing

\bibitem{bandwidth_plugin}
\BIBentryALTinterwordspacing
``Bandwidth plugin.'' [Online]. Available:
  \url{https://www.cni.dev/plugins/current/meta/bandwidth/}
\BIBentrySTDinterwordspacing

\bibitem{openshift}
\BIBentryALTinterwordspacing
``Openshift network plugins.'' [Online]. Available:
  \url{https://docs.openshift.com/container-platform/3.11/architecture/networking/network\_plugins.html}
\BIBentrySTDinterwordspacing

\bibitem{nbwguard}
\BIBentryALTinterwordspacing
C.~Xu, K.~Rajamani, and W.~Felter, ``Nbwguard: Realizing network qos for
  kubernetes,'' in \emph{Proceedings of the 19th International Middleware
  Conference Industry}, ser. Middleware '18.\hskip 1em plus 0.5em minus
  0.4em\relax New York, NY, USA: Association for Computing Machinery, 2018, p.
  32–38. [Online]. Available: \url{https://doi.org/10.1145/3284028.3284033}
\BIBentrySTDinterwordspacing

\bibitem{man7}
\BIBentryALTinterwordspacing
``Linux programmer's manual.'' [Online]. Available:
  \url{https://www.man7.org/linux/man-pages/man7/cgroups.7.html}
\BIBentrySTDinterwordspacing

\bibitem{sriov}
\BIBentryALTinterwordspacing
``Sr-iov network device plugin for kubernetes.'' [Online]. Available:
  \url{https://github.com/k8snetworkplumbingwg/sriov-network-device-plugin}
\BIBentrySTDinterwordspacing

\bibitem{fw_registry}
``firewall mark registry,'' \url{https://github.com/fwmark/registry}, 2020.

\bibitem{cni_spec}
\BIBentryALTinterwordspacing
``Container network interface (cni) specification.'' [Online]. Available:
  \url{https://www.cni.dev/docs/spec/}
\BIBentrySTDinterwordspacing

\bibitem{maplab}
T.~Schneider, M.~T. Dymczyk, M.~Fehr, K.~Egger, S.~Lynen, I.~Gilitschenski, and
  R.~Siegwart, ``maplab: An open framework for research in visual-inertial
  mapping and localization,'' \emph{IEEE Robotics and Automation Letters},
  2018.

\bibitem{Burri25012016}
\BIBentryALTinterwordspacing
M.~Burri, J.~Nikolic, P.~Gohl, T.~Schneider, J.~Rehder, S.~Omari, M.~W.
  Achtelik, and R.~Siegwart, ``The euroc micro aerial vehicle datasets,''
  \emph{The International Journal of Robotics Research}, 2016. [Online].
  Available:
  \url{http://ijr.sagepub.com/content/early/2016/01/21/0278364915620033.abstract}
\BIBentrySTDinterwordspacing

\bibitem{slam}
A.~Rensfelt, A.~C. Hernandez, B.~P. Gerö, C.~G. Blázquez, P.~C. Cubero,
  Y.~Nezami, and M.~Dohler, ``Network performance and the metaverse: Can 5g
  deliver what's needed?''
  \url{https://www.ericsson.com/en/blog/2022/11/network-performance-metaverse-5g},
  2022.

\end{thebibliography}
\end{document}